\documentclass[conference]{IEEEtran}

\usepackage[T1]{fontenc}
\usepackage{amsmath, amssymb}
\usepackage{graphicx}
\usepackage{algorithm2e}
\usepackage{enumerate}
\usepackage{multirow} 
\usepackage{url}
\usepackage{mathabx}
\usepackage{booktabs}
\usepackage{tabularx}
\usepackage{caption}
\usepackage{booktabs}
\usepackage{multirow}
\usepackage{siunitx}
\usepackage{arydshln}
\usepackage{color}
\usepackage{textcomp}
\usepackage{comment}

\newcommand{\ra}[1]{\renewcommand{\arraystretch}{#1}}

\begin{document}

%
%
\title{Population Estimation from\\ Mobile Network Traffic Metadata}

%
%
\author{
\IEEEauthorblockN{Ghazaleh~Khodabandelou\IEEEauthorrefmark{1}, Vincent~Gauthier\IEEEauthorrefmark{1}, Mounim~El-Yacoubi\IEEEauthorrefmark{1}, Marco~Fiore\IEEEauthorrefmark{2}}
\IEEEauthorblockA{\IEEEauthorrefmark{1}SAMOVAR, Telecom SudParis, CNRS, University Paris Saclay, France}
\IEEEauthorblockA{\IEEEauthorrefmark{2}CNR-IEIIT, Italy}
}

%
%
\IEEEoverridecommandlockouts
\IEEEpubid{\makebox[\columnwidth]{%
978-1-5090-2185-7/16/\$31.00~\copyright2016 IEEE \hfill}%
\hspace{\columnsep}\makebox[\columnwidth]{ }%
}

\maketitle

%
%
\begin{abstract}
Smartphones and other mobile devices are today pervasive across the globe. As an interesting side effect of the surge in mobile communications, mobile network operators can now easily collect a wealth of high-resolution data on the habits of large user populations. The information extracted from mobile network traffic data is very relevant in the context of population mapping: it provides a tool for the automatic and live estimation of population densities, overcoming the limitations of traditional data sources such as censuses and surveys. 
In this paper, we propose a new approach to infer population densities at urban scales, based on aggregated mobile network traffic metadata. Our approach allows estimating both static and dynamic populations, achieves a significant improvement in terms of accuracy with respect to state-of-the-art solutions in the literature, and is validated on different city scenarios.
\end{abstract}

%
%
\section{Introduction}
\label{sec:INTRODUCTION}
Large-scale datasets of mobile communication activity collected by network operators are becoming an important source of information for investigations of the spatiotemporal features of human dynamics. These datasets represent an important proxy of the movement patterns and social interactions of vast populations of millions of individuals, and allow investigating subscribers' endeavors at unprecedented scales.

Within this context, the study of population dynamics from mobile traffic data offers rich insights on human mobility laws~\cite{song2010limits}, disaster recovery~\cite{bagrow2011collective}, infective disease epidemics~\cite{bajardi2011human}, commuting patterns~\cite{yang14}, or urban planning~\cite{caceres12}. A comprehensive review is provided in~\cite{naboulsi16survey}. These studies have demonstrated how data collected by mobile network operators can effectively complement --or even replace-- traditional sources of demographic data, such as censuses and surveys. Indeed, classic data sources have major limitations in terms of compilation cost, scalability, and timely updating. Mobile traffic data can overcome all these issues, as it is relatively inexpensive to collect, covers large populations, and can be retrieved and analyzed in real-time or with minimum latency.

The features listed above are especially critical in the estimation of population densities. Extensive censuses of the populations living in urban and suburban regions are carried out at every few years by local authorities, at best. However, significant alterations of the static population distribution (\textit{i.e.}, the density of inhabitants based on their home locations) occur at shorter timescales, and cannot be tracked with conventional methods. Even more so when considering dynamic population distributions (\textit{i.e.}, the instantaneous density of inhabitants throughout, \textit{e.g.}, a day, based on their current position), which require repeated refreshing of the data during a same day.

Mobile network traffic data yields the potential to enable automated, near-real-time estimation of population density.
The correlation between the mobile communication activity and census data on static population density was first identified in~\cite{krings2009urban}. Since then, several works further investigated the possibility of exploiting telecommunication data, in isolation or mixed with other data sources, for the estimation of static and dynamic population distributions.
The current state-of-the-art methodology, presented in~\cite{douglass2015high}, leverages data from mobile traffic, map databases and satellite imagery to generate estimates of the static population with a correlation coefficient of 0.66 with ground-truth information.

In this paper, we build on the classical power relationship between mobile activity volume and population density~\cite{deville2014dynamic} and introduce a novel approach to the estimation of population. Tests with substantial real-world mobile traffic datasets prove that our model attains a significantly improved correlation with ground-truth data, typically in the range 0.80-0.87. Moreover:
\begin{itemize}
\item we design a solution based exclusively on metadata collected by the mobile network operator, avoiding complex and cumbersome data mixing;
\item we show that subscriber presence data inferred from the mobile communications of each user is a much better proxy of the population distribution than previously adopted metrics;
\item we introduce a number of original data filters that allow refining the population distribution estimates;
\item we evaluate our methodology in multiple urban scenarios, obtaining consistently good results;
\item we unveil the multivariate relationship between population density, subscriber presence and subscriber activity level;
\item we leverage our model to generate dynamic representations of the population distribution.
\end{itemize}

The document is organized as follows. Sec.\,\ref{sec:RELATEDWORK} reviews the literature related to our work. Sec.\,\ref{sec:DATASET} describes the reference original and derived datasets that we employ in our study. Sec.\,\ref{sec:NIGHTTIME} presents our proposed model for the estimation of static population distribution.
Sec.\,\ref{sec:DYNAMIC} shows model enhancements to deal with dynamic population densities.
Sec.\,\ref{sec:CONCLUSION} summarizes our work and draws conclusions.

%
%
\section{Related work}
\label{sec:RELATEDWORK}

There is wide agreement on the suitability of mobile network traffic data as a source of information for generic positioning analysis. Previous works have demonstrated that this type of data allows for the effective estimation of important places~\cite{csaji2013exploring}, the inference of trips among such places~\cite{bekhor2013evaluating}, and the derivation of origin-destination matrices by aggregating a large number of trips~\cite{calabrese2011estimating}.

As far as population distribution estimation is concerned, mobile communication data was first proposed as a proxy for the density of inhabitants in~\cite{ratti06}. Early evidences of the existence of an actual correlation between the mobile network activity and the underlying population density were presented in~\cite{krings2009urban}: the authors showed that city population sizes and the number of mobile customers follow similar distributions.

Subsequent works carried out more comprehensive evaluations. In~\cite{csaji2013exploring}, the home location of each subscriber was localized as the most frequently visited cell with a home profile (\textit{i.e.}, where the activity peak occurs in the evening). The density of home locations was then found to match very well --with a 0.92 correlation-- census data on nationwide population distribution. Similarly, excellent agreement between the overnight spatial density of mobile subscribers and that of nationwide static populations was found in~\cite{bekhor2013evaluating,calabrese2011estimating}. However, these results refer to a nationwide population, and the spatial granularity of the studies is counties or tracts (\textit{i.e.}, large regions comprising whole cities or macroscopic city neighborhoods).
Our focus is on intra-urban population distribution estimation: to that end, we downscale the study at the individual cell level, considering orders-of-magnitude higher accuracy and making the task much more challenging.

Citywide population estimation from mobile traffic data has been addressed by a limited number of works in the literature.
In~\cite{kang2012towards}, LandScan\texttrademark, a tool for ambient population estimation was employed to explore the relationship between the voice call activity and the underlying inhabitant density, at a 1-km$^2$ resolution. The authors found a weak correlation of 0.24, which improved to 0.45 by limiting the analysis to selected time intervals rather than considering the daily communication volume.
In~\cite{deville2014dynamic}, telecommunications data is mixed with a number of other sources, including information on Corine land use, OpenStreetMap infrastructure, satellite nightlights, and slope. This plethora of data is processed through a dasymetric modeling approach, resulting in high 0.92 correlation with census information. However, the correlation is a nationwide average, and the authors indicate that the accuracy is lower for the most densely populated areas, i.e., large cities. Indeed, in such areas, a normalized error of around 0.6 is measured in~\cite{deville2014dynamic}, whereas we obtain values below 0.06.

The approach in~\cite{douglass2015high} represents the current state-of-the-art in the estimation of cell-level population distribution from mobile traffic data. It performs random forest regression on the
median outgoing voice call volumes recorded on the whole urban region. This information is augmented with land use data extracted from OpenStreetMap (41\% of the spatial cells) and  satellite imagery (59\% of spatial cells).
We improve this approach in a number of ways. First, we use subscriber \textit{presence} metadata, which we prove to be a more sensible metric than the outgoing voice call volumes used in~\cite{douglass2015high}: in fact, this is a distinctive element of our study, telling it apart from all previous works in the literature. Second, we include filters based on daytime, land use information --which we infer from the mobile traffic activity itself-- and outlying human dynamics:
these allow us to account for important phenomena, such as the heterogeneity of subscribers' behaviors over the day and during the night, the diversity of mobile service usage in residential and non-residential area, or the variations of mobile traffic activity during weekdays and weekends or holidays. Third, we extend our evaluation to multiple cities, demonstrating the general viability of the methodology.
As a result, our model achieves a significant improvement in terms of estimation accuracy: we find a correlation of 0.83 in the Milan urban scenario, whereas the solution in~\cite{douglass2015high} yielded a 0.66 correlation in the same region.
Moreover, our solution relies on data collected from a single source, \textit{i.e.}, the mobile network operator, and does not need additional external data.

%
%
\section{Datasets}
\label{sec:DATASET}

In this work, we leverage several datasets made available by Telecom Italia Mobile (TIM) within their 2015 Big Data Challenge. 
Specifically, we focus on three major urban areas for which substantial data is available, i.e., the conurbations of Milan, Turin and Rome. For each city, we collect data describing the mobile traffic activity (Sec.\,\ref{sub:traffic}) and population distribution (Sec.\,\ref{sub:pop}), and we infer information on land use (Sec.\,\ref{sub:landuse}). 

\subsection{Mobile network traffic}
\label{sub:traffic}

The telecommunication data covers the months of March and April, 2015, and describes the volume of traffic divided by type (voice calls in/out, text SMS in/out, and Internet), and the presence of subscribers.
All metrics are aggregated in time and space. In time, the data is totaled over 15-minute time intervals. In space, metrics are computed over an irregular grid tessellation, whose geographical cells have sizes ranging from 255$\times$325 m$^2$ to 2$\times$2.5 km$^2$. The number of cells is 1419 for Milan, 571 for Turin and 927 for Rome.

\begin{figure}[tb]
  \centering
  \includegraphics[width=0.9\linewidth]{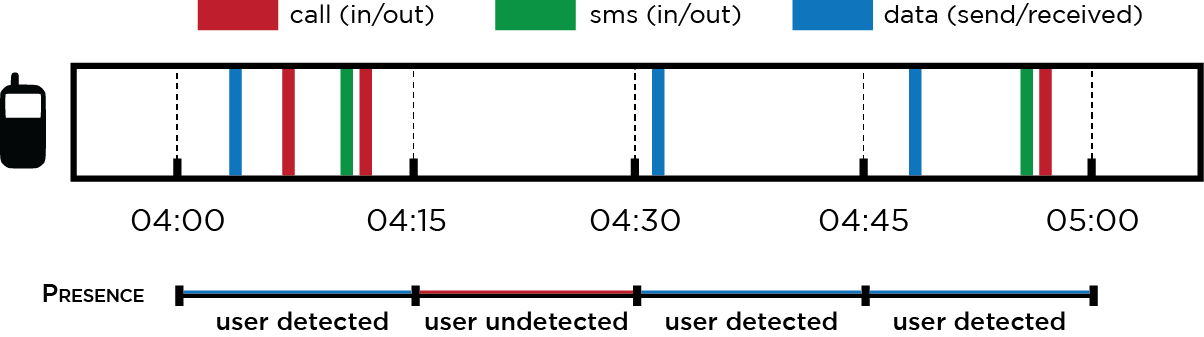}
  \caption{Subscriber presence in the TIM Big Data Challenge.}
  \label{fig:presence}
  \vspace*{-7pt}
\end{figure}

While voice, text, and data volumes are directly computed from the recorded demand,
the presence information is the result of a simple preprocessing performed by the mobile network operator. Basically, each subscriber is associated to the geographical cell where he/she  performed his/her last action (e.g., issued a call, received an SMS, and so on).
Therefore if the presence of a user is recorded in square $A$, then he performs an action in square $B$ at time $t_1$, and finally at time $t_2$ he interacts with the network in square $C$, her presence will be recorded as follows: for $t<t_1$, the presence of the user is recorded in $A$; at $t=t_1$, the presence of the user is moved from $A$ to $B$; for $t_1<t<t_2$, the presence of the user is recorded in $B$; at $t=t_2$, the user is moved from $B$ to $C$; for $t>t_2$, the presence of the user is recorded in $C$ until he performs an action in a different square. Fig.\,~\ref{fig:presence} shows an example of presence data preprocessing: the location of a specific user is detected every 15 minutes, if he entails at least one action during that time slot.

\subsection{Population distribution}
\label{sub:pop}

Population distribution data comes from the 2011 housing census in Italy by the national organization for statistics in Italy, ISTAT. It includes: population counts; a survey of structural attributes, update and review of municipal anagraphical lists; the number and structural features of houses and buildings. 
Specifically, population counts are measured in terms of families, cohabitants, persons temporarily present, domiciles, and other types of lodging and buildings, for each administrative area. 

In our study, we will consider such census data as a ground truth for the static population distribution in the reference urban regions.
To that end, we need to ensure spatial consistency between the mobile network traffic and the census data. We proceed as follows.
Let us denote as $U_j$ the total number of inhabitants in the administrative area $j$, and as $A_j$ its surface. The population density $\rho_i$ in a geographical cell $i$ defined in Sec.\,\ref{sub:traffic} is then computed as
\begin{equation}
\rho_i=\frac{1}{A_i}\sum_{j=1}^K U_j \frac{A_{i \cap j}}{A_j}
\label{eq:density}
\end{equation}
where $A_i$ is the surface of cell $i$, $K$ denotes the total number of administrative areas, and $A_{i \cap j}$ stands for the intersection surface of cell $i$ and administrative area $j$.

\subsection{Land use}
\label{sub:landuse}

Land use information is critical to the accurate estimation of population densities, as already demonstrated in the recent literature~\cite{douglass2015high}.
In this work, we leverage the telecommunication data itself to classify the geographical cells based on their primary land use. To that end, we employ MWS, which is the current state-of-the-art technique for land use detection from mobile traffic data~\cite{furno2015comparative}.
MWS computes, for each spatial cell of the target region, a mobile traffic signature, i.e., a compact representation of the typical dynamics of mobile communications in the considered cell. Specifically, MWS signatures are computed as the median voice call and texting activity in a cell recorded at every hour of the week. Signatures are then clustered based on their shape. This allows inferring a limited set of archetypal signatures, which are representative of distinct types of human activities. When applied to our reference datasets, MWS identifies seven major activity types: residential, office, transportation, touristic, university, shopping and nightlife. Clearly, all cells whose signatures are clustered into, e.g., the residential archetypal signature, can be classified as residential areas. Overall, this allows identifying land uses in our reference region.

It is important to stress that this approach is sensibly different from those used in previous works addressing population density estimation. In fact, we do not rely on external data sources such as, \textit{e.g.}, crowdsourced map databases or post-processed satellite imagery as, \textit{e.g.}, in~\cite{douglass2015high}. Instead, we resort solely to data collected by the mobile network operator. On the one hand, this makes the solution presented in this paper \textit{self-contained}, \textit{i.e.}, practicable by using just one type of dataset and without any need for dataset mixing -- whose blend of data of different natures is often cumbersome. On the other hand, land uses detected through mobile traffic data can be much more accurate and meaningful than those returned by other methods. As an example, the Milan conurbation territory is classified into buildings, vegetation, water, road, and railroads in~\cite{douglass2015high}: these categories are based on pure geographical features, and have small relation with the activity of individuals. Instead, the classes identified by MWS in the same region, listed above, map to a diversity of actual human endeavors, and, as such, have a stronger tie to the population density we aim at estimating. Our results will corroborate this intuition.

%
%
\section{Static population estimation}
\label{sec:NIGHTTIME}

The baseline population estimation model we adopt builds on previous results that demonstrated a consistent power relationship between the mobile network traffic activity $\sigma_i$ and the population density
$\rho_i$ at a same region $i$~\cite{douglass2015high, deville2014dynamic}. Thus
\begin{equation}
\rho_i =\alpha \sigma_i^\beta,
\label{eq:regression}
\end{equation}
where the parameters of $\alpha$ and $\beta$ represent the intercept (or the scale ratio) and slope (or the effect of $\rho_i$ on $\sigma_i$) of the model, respectively. By transforming the formula to a logarithmic scale, we obtain $log(\rho_i) = log(\alpha) + \beta log(\sigma_i)$. We can then use a regression model to estimate the parameters $\alpha$ and $\beta$ in the expression~(\ref{eq:regression}).

\begin{figure}[tb]
  \centering
  \includegraphics[width=0.67\linewidth]{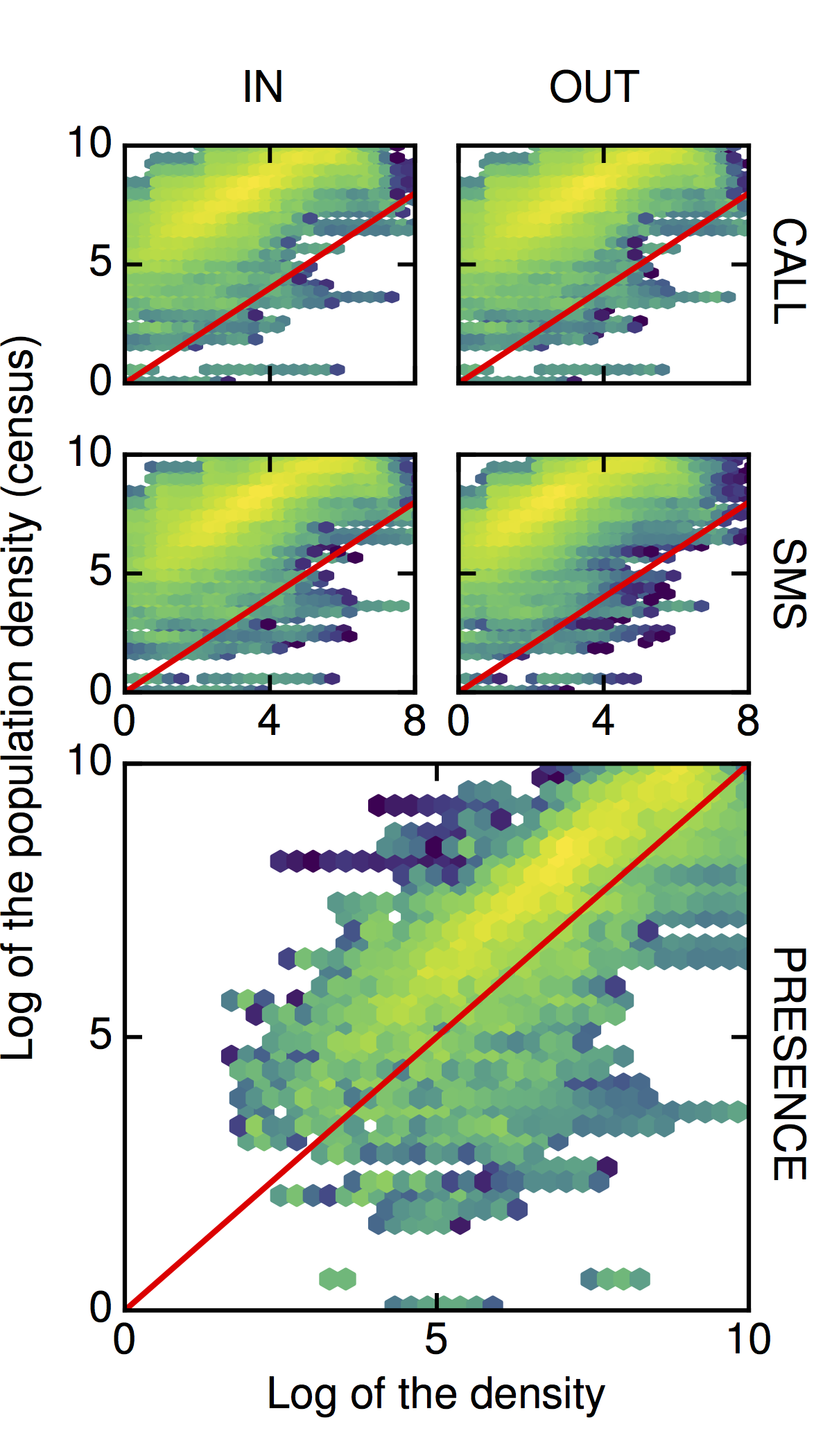}
  \caption{Milan. ISTAT population density as function of the call volume, SMS volume, and subscriber presence.}
  \label{fig:hetero}
  \vspace*{-5pt}
\end{figure}

Unfortunately, the regression yields poor results if run on the raw telecommunication data, no matter the data type. Let us consider the data for Milan: the simple visualization in Fig.\,\ref{fig:hetero} unveils that the datasets described in Sec.\,\ref{sub:traffic} suffer from heterogeneity and heteroscedasticity. The plots illustrate the density of calls, SMS and presence\footnote{We omit Internet traffic data as it results in even sparser density maps. Indeed, it is widely acknowledged that mobile Internet data, affected by traffic generated by applications running in background, is less representative of human activity than voice and texting~\cite{furno2015comparative}.}
as a function of the associated population density in Sec.\,\ref{sub:pop}. The high variance between dependent and independent variables reveals the absence of homoscedasticity in the data. However, the classical regression models assume that there is no heteroscedasticity in the data, and, based on this assumption, infer the best linear unbiased estimators along with the lowest variance among all other unbiased estimators.

It follows that filtering the noisy data in Fig.\,\ref{fig:hetero} is a necessary step toward applying effective regression models. Next, we discuss data de-noising on multiple levels. We focus on the Milan case study, and then generalize the results of our approach by testing it in other city scenarios.


\subsection{Data type}
\label{sec:variablesfiltering}

\begin{figure}[tb]
  \centering
  \includegraphics[width=1.0\linewidth]{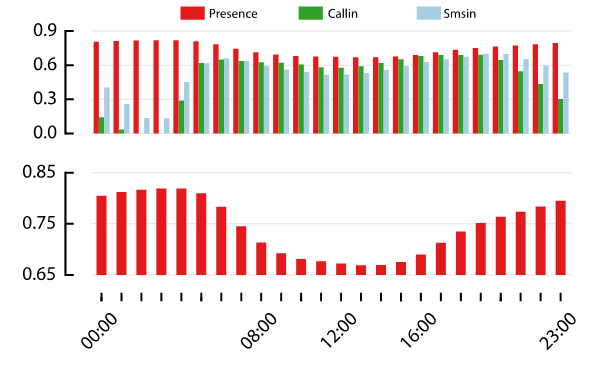}
  \caption{Milan. Top: correlation between different types of mobile network traffic data and the population census density, on a hourly basis. Bottom: zoom on presence metadata.}
  \label{fig:type} 
  \vspace*{-9pt}
\end{figure}

The very first step we take is determining which type of mobile network traffic metadata is the most reliable for population estimation.
To that end, we study the correlation (i.e., the Pearson correlation coefficient) between the population census data and the different metrics presented in Sec.\,\ref{sub:traffic}, as recorded on a per-cell basis.

The top plot in Fig.\,\ref{fig:type} shows the correlation coefficient value obtained for incoming voice calls and text messages%
\footnote{Equivalent results were obtained for outgoing voice calls and text messages, and are not shown for the sake of brevity.},
and subscriber presence. The correlation is separated on a hourly basis to evidence the impact of daytime.
Our results are aligned with those in~\cite{douglass2015high}, which, however, only considered calls and SMS, and not the subscriber presence. As a result, their conclusion was that calls between 10 am and 11 am yield the strongest correlation with population density.

In fact, Fig.\,\ref{fig:type} shows that subscriber presence is a much fitter metric in that sense. The correlation values for presence metadata are consistently higher than those for voice calls and text SMS. Thus, our choice of data type is subscriber presence. We stress that we are the first to consider this type of mobile network traffic metadata for population estimation, which is often undisclosed by mobile operators for privacy reasons.


\subsection{Hour filtering}
\label{sec:timefiltering}

A second dimension on which to perform data filtering is the temporal one. As found in the literature~\cite{deville2014dynamic}, the correlation between mobile network traffic data and the population density varies over time. This can be verified in the top plot of Fig.\,\ref{fig:type}.

In fact, an even clearer picture of the phenomenon is provided by the zoomed bottom plot, which focuses on presence metadata.
The highest correlation coefficient is recorded at night, between 4 am and 5 am. We underscore that this is a very reasonable outcome, since the ISTAT population data refers to housing, and individuals are most likely at home overnight. 
Thus, we pick subscriber presence during that specific hour as the mobile network traffic metadata for the regression model. Henceforth, $\sigma_i$ defined in~(\ref{eq:density}) describes the user presence in cell $i$ between 4 am and 5 am.


\subsection{Day filtering}
\label{sec:daysfiltering}

\begin{figure*}[!ht] 
   \centering
   \includegraphics[width=0.8\linewidth]{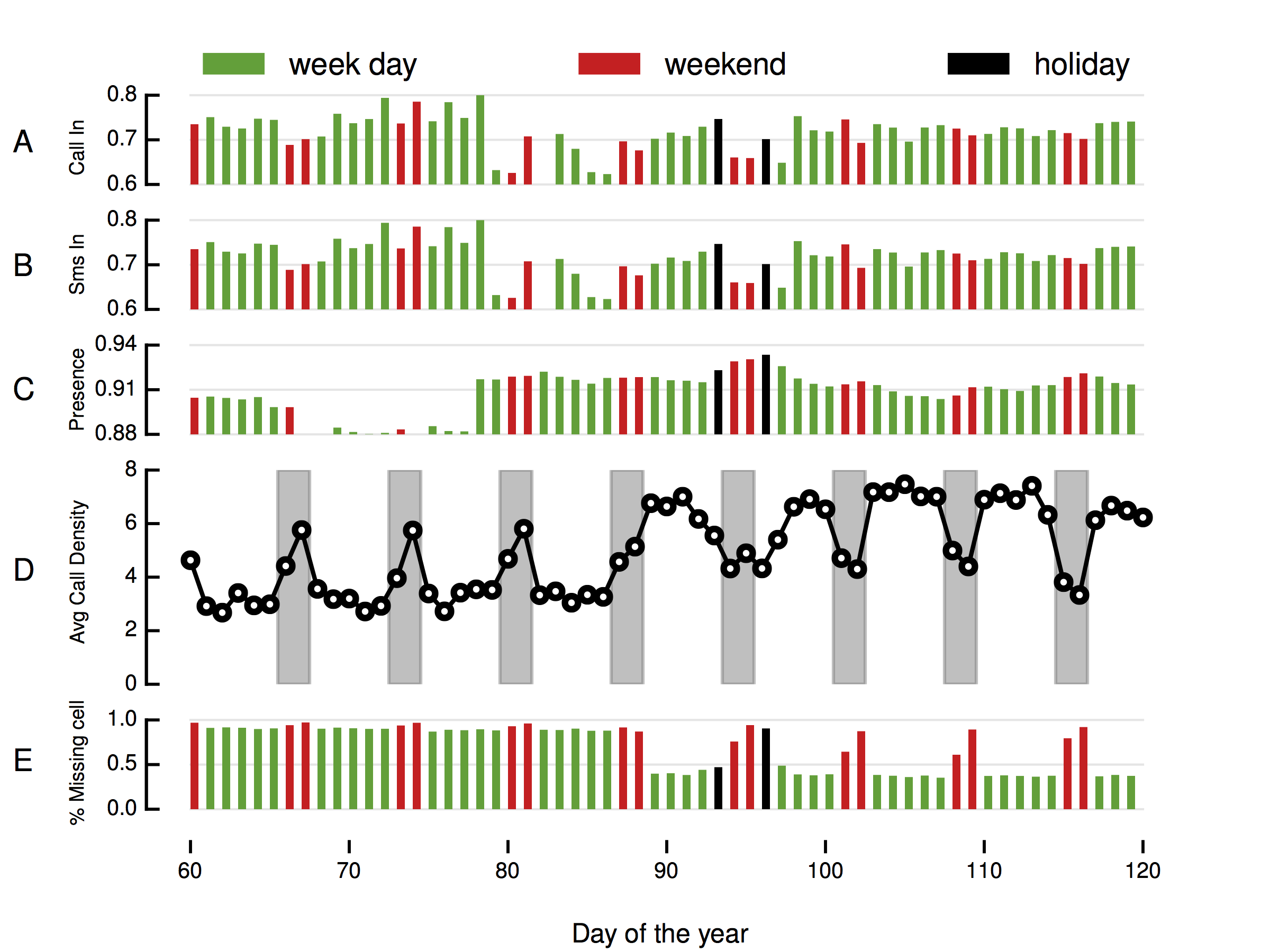}
   \caption{Milan. Pearson coefficient of correlation for different mobile network traffic data, in March and April 2015, between 4 am and 5 am. A) Incoming call (range up to 0.8), B) Incoming SMS (range up to 0.8), C) Subscriber presence (range starts at 0.88). D) Average call density, with weekends highlighted in gray. E) Percentage of cells for which no information is available.}
   \label{fig:day}
   \vspace*{-14pt}
\end{figure*}

As mentioned in Sec.\,\ref{sub:traffic}, the mobile network traffic data we employ covers 60 days in March and April, 2015. An interesting question is if all these days should be considered in the regression model, or if there exists a more meaningful subset of days. Indeed, mobile network traffic data is clearly affected by the diverse activity patterns and social phenomena that may characterize different days.

The three top plots in Fig.\,\ref{fig:day} show the Pearson coefficient of correlation over the 60 days of our reference datasets. They refer to incoming voice calls (A), incoming SMS (B), and presence (C), and, consistently with the previous results, are computed over the hourly interval 4-5 am.
The plots confirm our previous conclusion on the relevance of subscriber presence as a proxy of population density: the correlation coefficient ranges between 0.88 and 0.94, i.e., much higher values than those attained by calls and SMS, in the range between 0.6 and 0.8.
No clear trend of the correlation coefficient is instead observed over the 60 days, for all data types.

A more insightful result is obtained by considering the average call density, i.e., the average number of calls (per km$^2$) recorded during a 15-minute slot in a spatial cell, between 4 am and 5 am (D). In this case, a remarkable weekly pattern appears, with weekdays and weekends (highlighted in gray) resulting in different behaviors. This recurrent phenomenon is an indicator that weekends may yeild different behaviors in terms of population presence (e.g., due to much increased nightlife), which could introduce a significant bias in our study. Thus, we filter out weekends from our data.
Further motivation for this choice comes from the bottom plot in Fig.\,\ref{fig:day}, showing the percentage of cells for which no data is available between 4 am and 5 am on each day (E). We remark that there exists a high percentage of missing data during weekends (denoted by red bars) and holidays (denoted by black bars, and corresponding to Good Friday and Easter): we thus filter out holidays as well from our data.

Finally, plots D and E in Fig.\,\ref{fig:day} also highlight an evident diversity between the month of March and April. We speculate that this is due to issues in the data collection process.


\subsection{Regression with RANSAC}
\label{sec:RANSACRegression}

\begin{figure}[ht]
  \centering
  \includegraphics[width=0.7\linewidth]{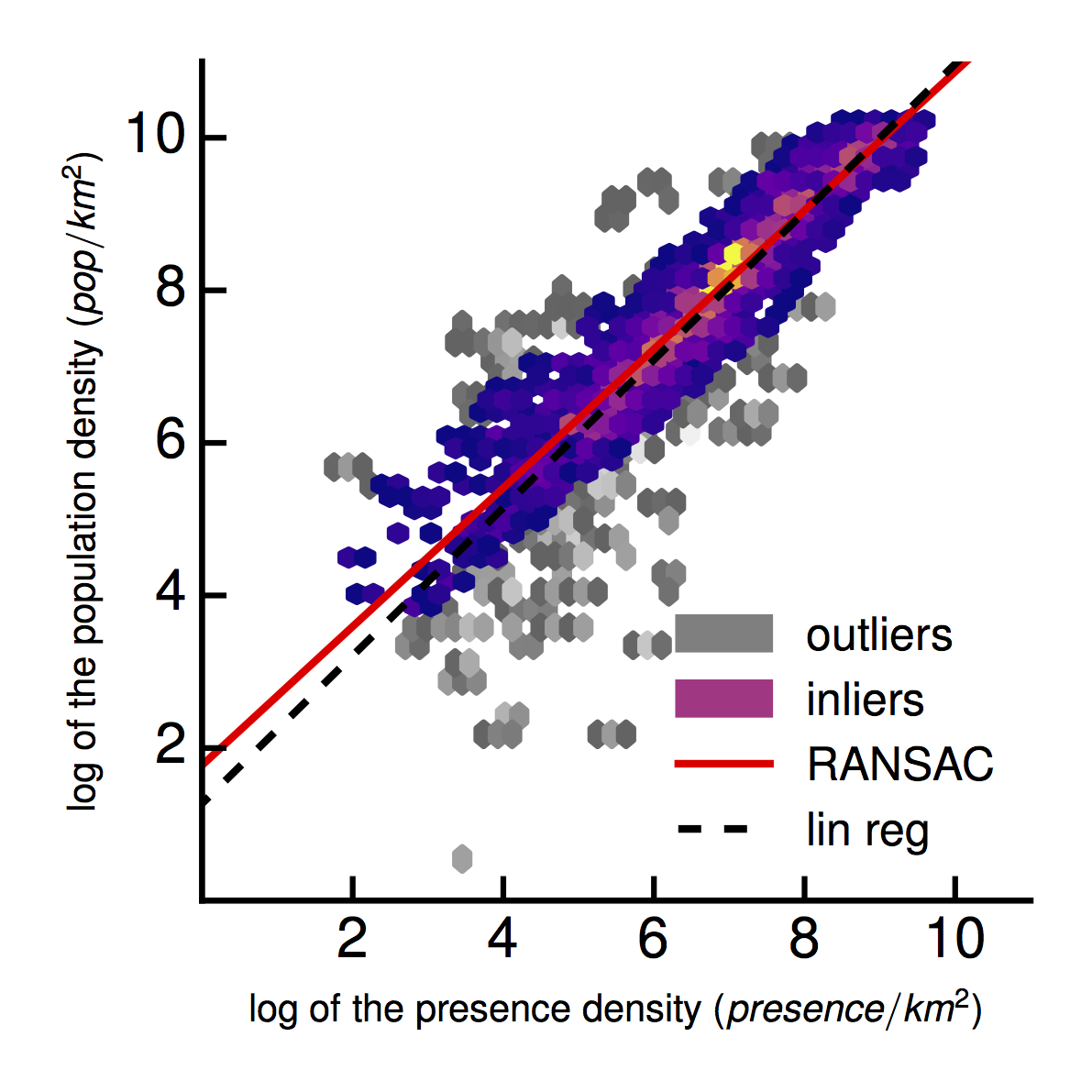}
  \caption{Milan. RANSAC and linear regression on filtered subscriber presence data. Figure best viewed in color.}
  \label{fig:ransac} 
  \vspace*{-14pt}
\end{figure}

In order to estimate the parameters $\alpha$ and $\beta$ in (\ref{eq:regression}), we employ the RANSAC regressor
on the filtered presence data. RANSAC estimates the parameters of a model from observations in an iterative manner, and automatically detects and excludes outlying points. Fig.\,\ref{fig:ransac} illustrates the fitting with the RANSAC regression model, in a log-log scale (solid red line).
The inliers and outliers detected in the data by RANSAC are shown in violet and gray, respectively. The estimated parameters are $\hat{\alpha}= 1.265$, $\hat{\beta}=0.979$.
The result underscores the quasi-linearity of the relation between subscriber presence and population, as also evidenced by the decent match with a linear fitting (dashed black line).

\begin{figure}[tb]
  \centering
  \includegraphics[width=0.95\linewidth]{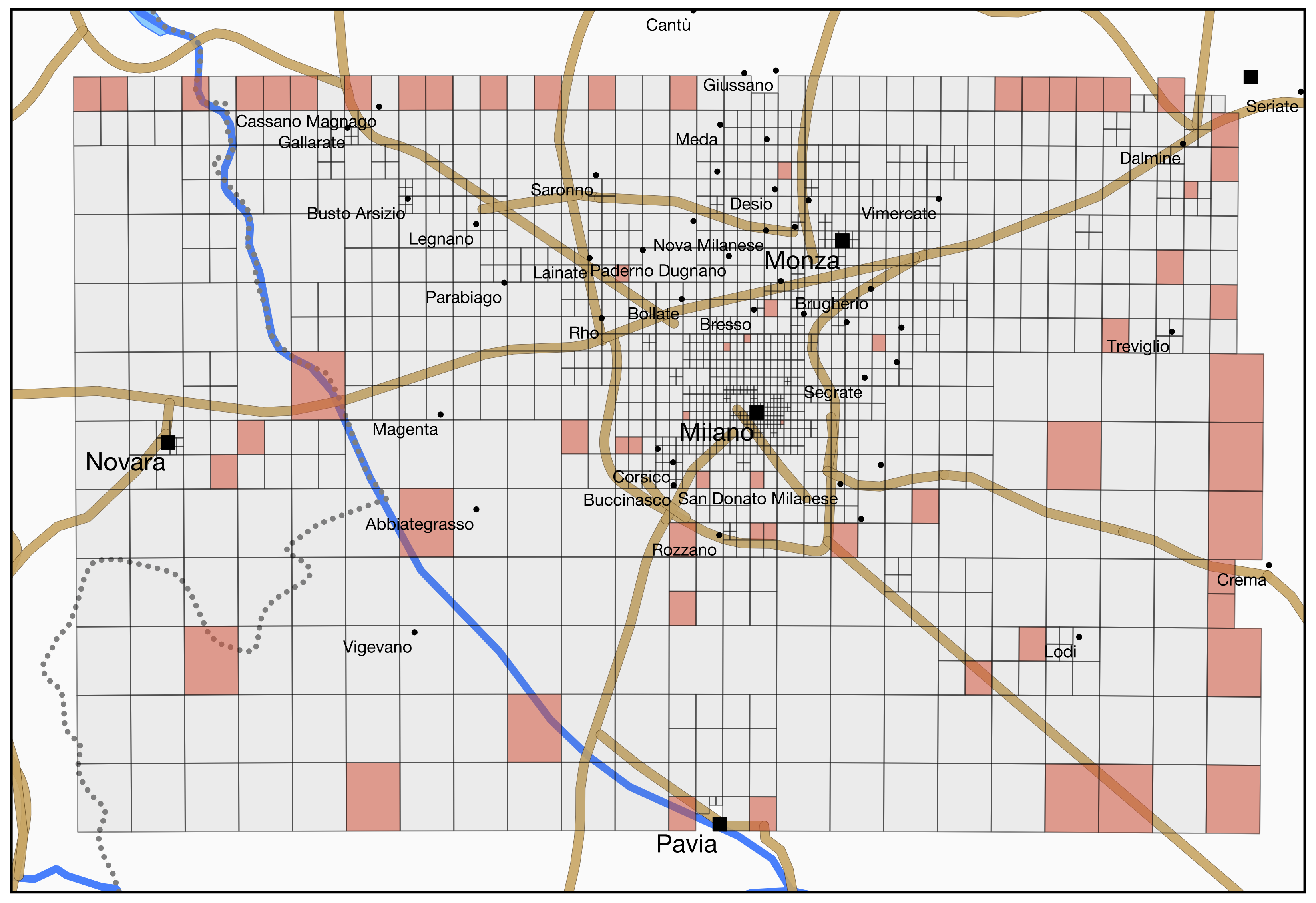}
  \caption{Milan. Geographical distribution of the cells that most frequently entail outliers detected by RANSAC.}
  \label{fig7:outliers}
  \vspace*{-7pt}
\end{figure}

It is interesting to have a closer look at the outliers detected by RANSAC. As a matter of fact, an analysis of these points reveals that outliers refer to a subset of cells, consistently over time. Thus, those cells yield some features that make their associated subscriber presence less related to their local population density.
A map of such cells is portrayed in Fig.\,\ref{fig7:outliers}.
It is evident that most outlying behaviors are encountered at the borders of the region. We speculate that this is an artifact of the spatial tessellation used in our reference datasets: clearly, the tessellation is a rough approximation of the cellular coverage, and it may well be that some border effect emerges, with an artificial (too high or too low) mobile network traffic associated to border cells.

We also observe cells with outlying behaviors within the geographical region under consideration in Fig.\,\ref{fig7:outliers}. We hypothesize that these are areas where the population density changed significantly over the last few years. As a matter of fact, the ISTAT census dates back to 2011, whereas the mobile network traffic data refers to 2015. It is likely that the distribution of the local population has undergone some evolution during the four years that separate the two datasets.
For instance, a confirmation to this consideration seems to come from the high density of outlying cells in the Southern periphery of Milan (in proximity of the South beltway): this is an expansion area of the city, and the mobile network traffic activity reflects a density of inhabitants in 2015 that is significantly higher than that recorded in 2011.
We deem this result to be an evidence of how mobile network traffic data could help updating population distribution maps, without any cost and with very limited latency.


\subsection{Model evaluation}
\label{assess}

The regression model allows computing an estimate of the static population $\hat{\rho}_i$ from the presence metadata $\sigma_i$, within each spatial cell $i$, as 
\begin{equation}
\hat{\rho}_i =\hat{\alpha} \sigma_i^{\hat{\beta}}.
\label{eq:estimreg}
\end{equation}

The estimated population $\hat{\rho}_i$ can then be compared to the ground-truth data $\rho_i$ from the ISTAT census.
To that end, we use the determination coefficient $R^2$, and the Normalized Root Mean Square Error ($NRMSE$). The former provides a measure of the quality of the fitting of the estimates on ground-truth data.
The latter describes the fraction of the error between values predicted by the model and the values of census population data.
The $R^2$ coefficient is computed as
\begin{equation}\label{eq:r2}
R^2 = 1-\frac{\sum_{i=1}^N \left(\rho_i - \hat{\rho}_i\right)^2}{\sum_{i=1}^N \left(\rho_i-\bar{\rho}\right)^2},
\end{equation}
where $N$ denotes the number of cells in the spatial tessellation, and $\bar{\rho}$ is the overall average population density computed on all cells.
The $NRMSE$ facilitates the comparison of the model results in different contexts. It is defined as
\begin{equation}\label{eq:RMSE}
NRMSE = \frac{1}{\rho_{max}-\rho_{min}}\sqrt{\frac{\sum_{i=1}^N(\hat{\rho}_i-\rho_i)^2}{N}},
\end{equation}
where $\rho_{max}$ and $\rho_{min}$ are the maximum and minimum population densities recorded in the target region.

Since the model is trained on ISTAT census data, we adopt a two-fold cross-validation procedure, as follows.
For each test, we separate the ground-truth data into two subsets: two-thirds of the data are used as a training set, and the remaining one-third as a test set. Then, the training set is used to learn the model parameters, and the resulting model is evaluated against the test set.

\begin{figure}[tb]
  \centering
  \includegraphics[width=1\linewidth]{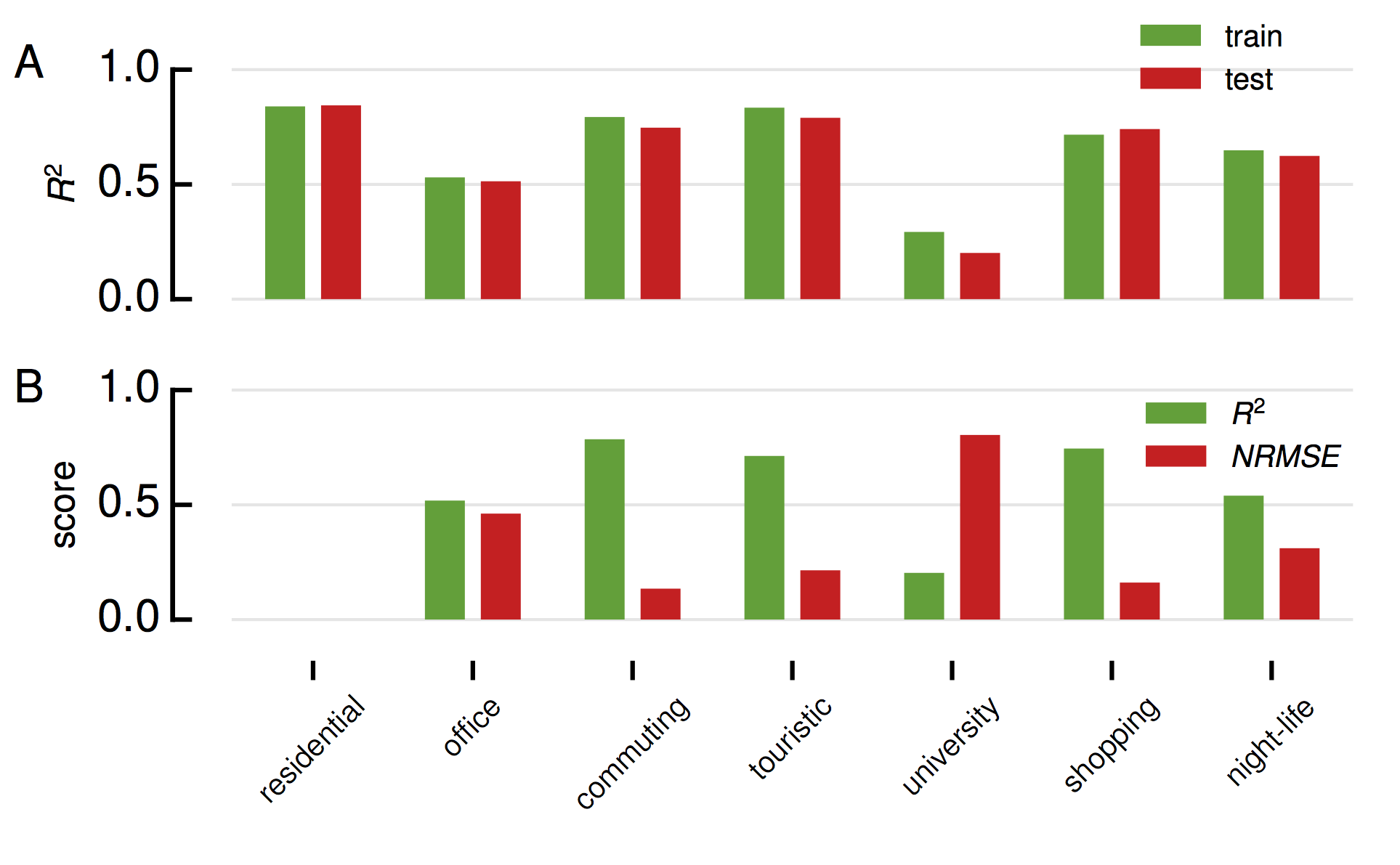}
  \caption{Milan. Model evaluation. A) $R^2$ for training and test sets, separated by land use. B) $R^2$ and $NRMSE$ of the model trained on residential land use on the other land uses.}
  \label{fig:corrclus} 
  \vspace*{-7pt}
\end{figure}

The baseline result, in the Milan case study, is shown in Fig.\,\ref{fig:corrclus}. The top plot (A) shows the determination coefficient $R^2$ obtained with the training and test data. A first observation is that results are comparable for training and test data, which validates our model.

A second remark is that the results are separated by land use, an information that we extract from the mobile network traffic data itself, as detailed in Sec.\,\ref{sub:landuse}.
The rationale for this choice is that land uses affect the activity of individuals, possibly including their mobile communication habits. For instance, the mobile network traffic generated in a residential zone undergoes dynamics that are very different from those observed in a shopping area. In turn, this diversity can induce variations in the quality of the population estimation model across land uses. Plot (A) in Fig.\,\ref{fig:corrclus} shows precisely this effect: some land uses have higher $R^2$ than others. Specifically, residential zones display the highest fitting score, with $R^2=0.84$ and a $NRMSE$ of 0.046 (not shown). This is reasonable, given that in these areas the majority of the communication activity is generated by users in their homes.

We also verified up to which extent a model trained on residential land use can be used to estimate populations in areas characterized by different land uses.
The bottom plot (B) in Fig.\,\ref{fig:corrclus} shows the results in the Milan case study. Performance are good for commuting, touristic, and shopping zones, fair for office and nightlife zones, and bad for university areas.
We speculate that this phenomenon could be due to the presence of overnight mobile communication activities in university campuses (e.g., parties or measurement for research purposes) where nobody actually lives.
In all cases, university areas only represent a negligible minority of spatial cells, which allows concluding that a model trained on residential land use can be effectively used to estimate the population density in a whole urban region.


\subsection{Other city use cases}
\label{sub:cities}

\begin{table*}[tb]
\caption{$R^2$ and $NRMSE$ in different cities of Italy, for residential-only and mixed land use.\vspace*{-4pt}}
\centering
\ra{1.3}
\begin{tabular}{@{}lllllllllllllll@{}}
& \multicolumn{10}{c}{\textsc{Residential}} & \phantom{abc} & \multicolumn{2}{c}{\textsc{Mixed}} \\
\cmidrule{2-11} \cmidrule{13-14}
&& \multicolumn{5}{c}{Training}  && \multicolumn{3}{c}{Test} &&& \\
\cmidrule{2-8} \cmidrule{10-11}
& ${\scriptstyle \hat{\alpha}}$ & ${\scriptstyle 95\% C.I.}$ & ${\scriptstyle \hat{\beta}}$ & ${\scriptstyle 95\% C.I.}$& &${\scriptstyle R^2}$ & ${\scriptstyle NRMSE}$ & &${\scriptstyle R^2}$ & ${\scriptstyle NRMSE}$ && ${\scriptstyle \bar{R^2}}$ & ${\scriptstyle \overline{NRMSE}}$ \\
\midrule
\textsc{Milan} &1.24&[1.03,1.37]&0.97 &[0.89,1.0]&&0.84 & 0.046 && 0.83 & 0.047 && 0.80 & 0.059 \\
\cdashline{1-14}[2pt/3pt]
\textsc{Turin} &0.94&[0.82,1.2]&0.99&[0.88,1.2]&& 0.80 & 0.052 && 0.80 & 0.053 &&0.76&0.065   \\
\cdashline{1-14}[2pt/3pt]
\textsc{Rome}  & 0.75&[0.67,0.98]&1.03&[0.92,1.1]&& 0.87 & 0.035 && 0.87 & 0.035 &&0.84 &0.044  \\
\bottomrule  
\label{tab:cities}
\vspace*{-23pt}
\end{tabular}
\end{table*}

All previous results refer to the Milan case study. We generalize our analysis by considering two other major cities in Italy, i.e., Rome and Turin.

For each urban case study, we adopt the aforementioned cross-validation procedure, separating training and test datasets. We then estimate the model parameters $\hat{\alpha}$ and $\hat{\beta}$: consistently with our previous results, this is performed by considering training data related to residential land use only. 
The {\it Residential} portion of Tab.\,\ref{tab:cities} shows the results we obtain, in terms of parametrization of $\hat{\alpha}$ and $\hat{\beta}$ and quality of the estimation on the test set of remaining residential areas. 
The right portion of the Tab.\,\ref{tab:cities}, denoted as {\it Mixed}, refers to the quality of the estimation on all areas, including those that are not residential in nature: it thus gives us information on the accuracy of a model trained on residential data only, when used on a complete urban region.
In this second case, we employ the following averaged quality metrics:
\begin{equation}
\bar{R^2}=\sum^L_{\ell=1} \frac{N_\ell}{N} R_\ell^2, 
\end{equation}
\begin{equation}
\overline{NRMSE}=\sum^L_{\ell=1} \frac{n_\ell}{N} NRMSE_\ell,
\end{equation}
where $L$ is the number of different land uses, $N_\ell$ stands for number of spatial cells associated to a given land-use $\ell$, and $R_\ell^2$ (respectively, $NRMSE_\ell$) are the determination coefficient (respectively, normalized root mean square error) computed on land use $\ell$. Thus, these metrics provide an weighted average of the performance of the estimation over all land uses.

The results in Tab.\,\ref{tab:cities} are fairly close for all cities. Rome shows highest scores, with $R^2$=0.87 and $NRMSE$=0.035 for residential land use, and $R^2$=0.84 and $NRMSE$=0.044 for the overall case. Milan is a close second, and Turin performs slightly worse. Yet, the $R^2$ we measure is consistently higher than that observed in recent previous works, e.g.,~\cite{douglass2015high}.

\begin{figure}[tb]
  \centering
  \includegraphics[width=0.95\linewidth]{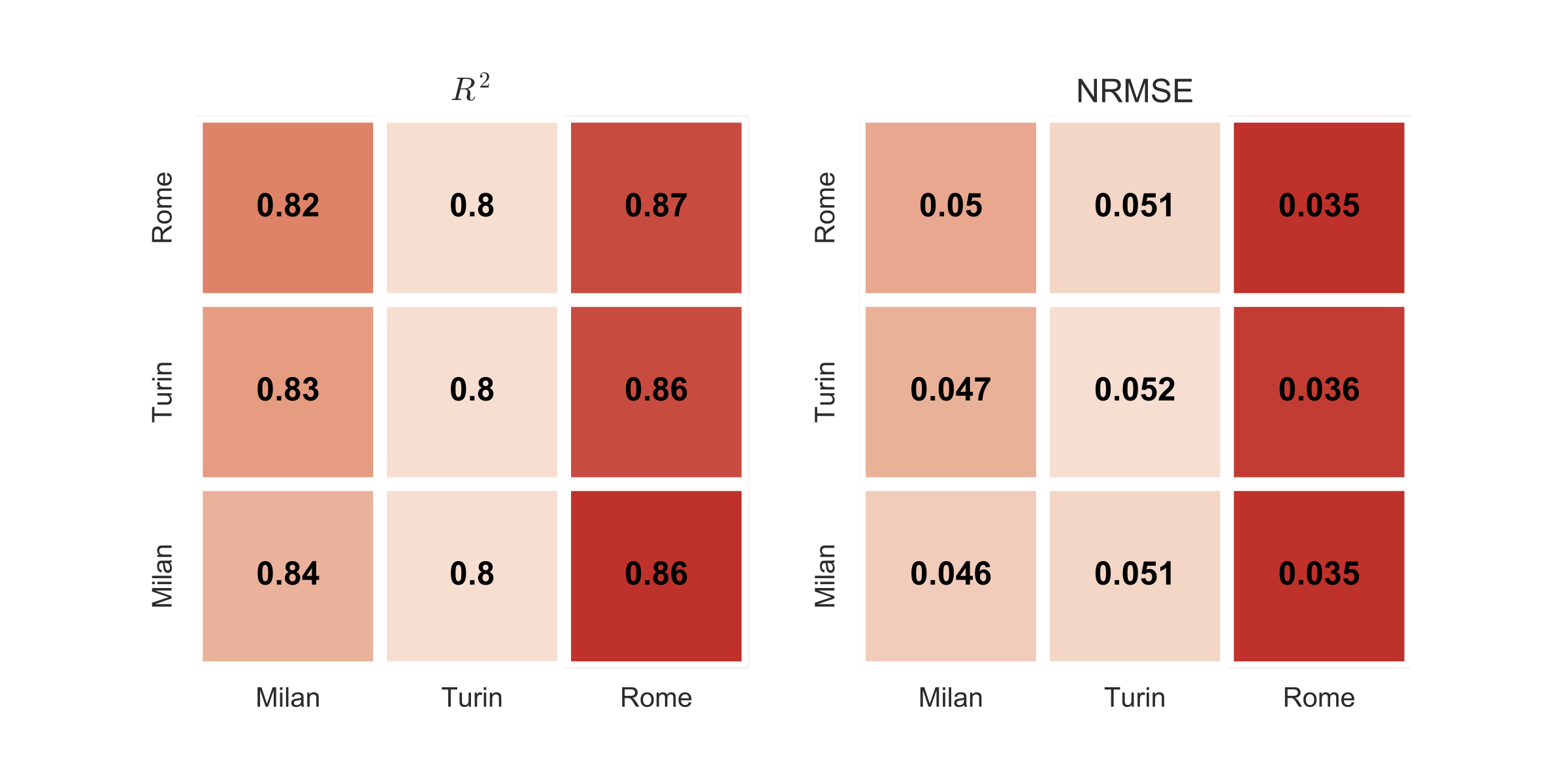}
  \caption{Cross-city test: models trained on data collected in cities along rows are used to estimate the population of cities along columns. Tables refer to $R^2$ and $NRMSE$.}
  \label{fig:corr_mat} 
  \vspace*{-14pt}
\end{figure}

From Tab.\,\ref{tab:cities}, we observe that the parameters $\hat{\alpha}$ and $\hat{\beta}$ are not dramatically different across cities. We thus explore the possibility of estimating the population in an urban area using a model trained on data collected in another city.
Fig.\,\ref{fig:corr_mat} summarizes our findings.
Overall, we find high $R^2$ and low $NRMSE$ in all cases, which let us conclude that a cross-city estimation of population is possible.
This is an important observation, paving the road to the estimation of populations in cities for which mobile network traffic data is available, but where no ground-truth population distribution is provided.

%
%
\section{Dynamic population estimation}
\label{sec:DYNAMIC}

In the previous discussion, we have focused on static populations, considering the distribution of inhabitants based on their home locations.
However, an interesting possibility offered by mobile network traffic analysis is that of estimating dynamic populations, i.e., the variations in the distribution of inhabitants determined by their daily activities.
Clearly, this is an even more challenging task, due to the faster timescale at which such daily dynamics occur.

The main problem one has to cope with in estimating dynamic populations is the lack of ground truth data, which makes it impossible training a model such as that in (\ref{eq:estimreg}). At the same time, reusing the parameters $\hat{\alpha}$ and $\hat{\beta}$ computed for the static population may be risky, because there is no certainty that the relationship between the subscriber presence $\sigma_i$ and the static population remains valid for dynamic populations.

Our approach consists in drawing a novel multivariate relationship between the population distribution, the subscriber presence and the level of mobile communication activity of subscribers. This allows taking the constants $\alpha$ and $\beta$ out of the equation, and finding a unifying equation that can be used to estimate dynamic populations.

\subsection{Subscriber presence and activity level}
\label{sub:dyn-activity}

We start by discussing the interplay between the presence and the mobile communication activity level.
The latter is formally defined as the frequency with which a subscriber interacts with the mobile network.

\begin{figure}[tb]   
    \centering
    \includegraphics[width=0.95\linewidth]{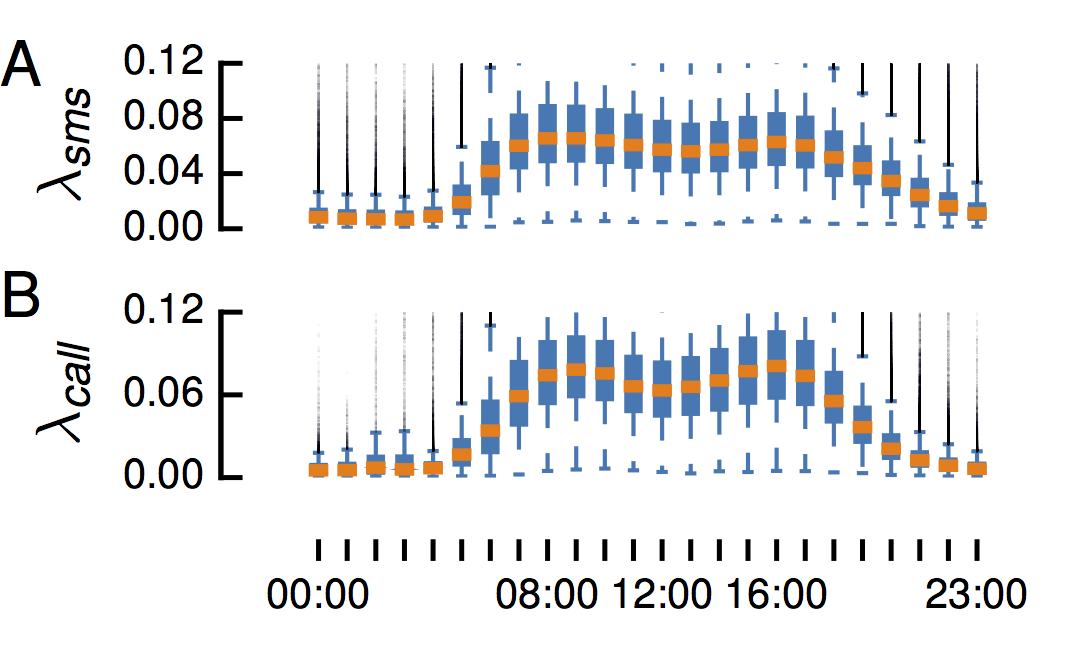}\vspace*{-4pt}
    \caption{Subscriber activity level for voice calls (A, top) and SMS (B, bottom) over daytime.\vspace*{-3pt}}
    \label{fig2:activ}    
    \vspace*{-10pt}
\end{figure}

\begin{figure}[tb] 
	\centering
	\includegraphics[width=0.8\linewidth]{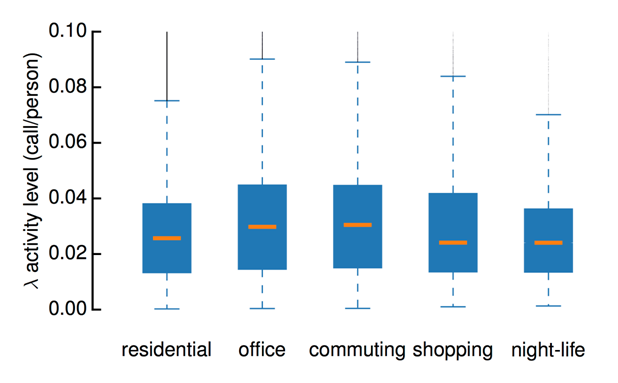}
    \caption{Subscriber activity level for calls, split by land use.\vspace*{-2pt}}
    \label{fig3:activlanduse} 
    \vspace*{-10pt}
\end{figure} 

Fig.\,\ref{fig2:activ} depicts the average level of activity per subscriber, as recorded in the data over the 24 hours of each day. It is expressed as the mean number of events per user ($\lambda$) observed in the data at every 60 minutes.
We remark a significant variation of activity, with minimum network usage at night and increased mobile communications during the working hours: this is consistent with previous analyses of mobile traffic dynamics.
Interestingly, differences across land uses are moderate, as shown on Fig.\,\ref{fig3:activlanduse}.
We conclude that the mobile communication activity is heterogeneous, and such a behavior emerges in time more than over land uses.

Such heterogeneity in the activity level has an impact on the correctness of the presence information. Let us consider again Fig.\,\ref{fig:presence}: the more often a mobile device sends or receives calls, SMS, and data packets, the more accurate is its localization in the presence dataset.
A legitimate question is then if the heterogeneous activity we discussed can be linked to the model parametrization, and explain -- in part or in full -- the diversity of values of $\hat{\alpha}$ and $\hat{\beta}$ observed in Sec.\,\ref{sec:NIGHTTIME}.

\subsection{Population estimation with activity level}
\label{sub:dyn-estimation}

\begin{figure}[tb]   
    \centering
    \includegraphics[width=1\linewidth]{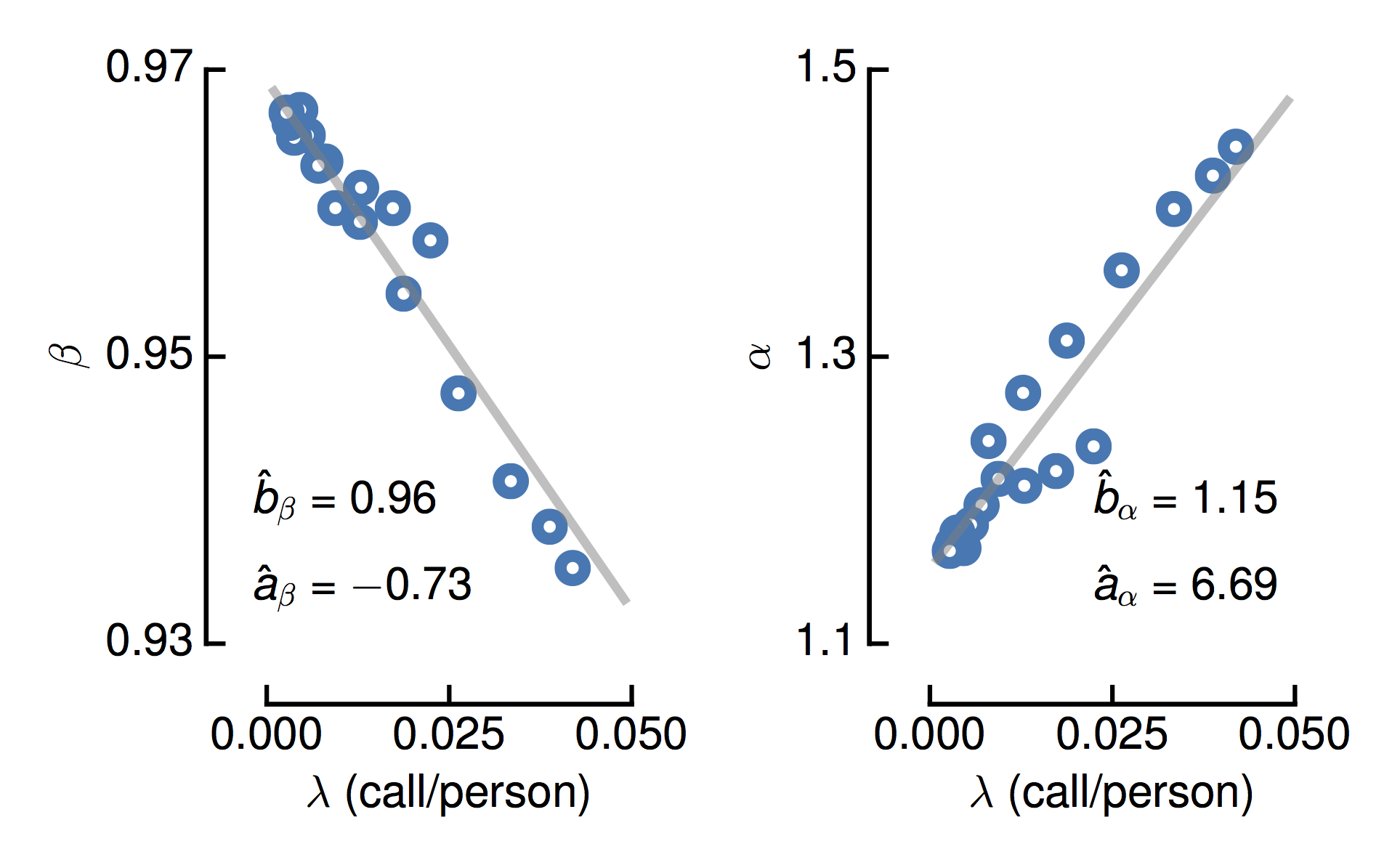}\vspace*{-4pt}
    \caption{Linear relationship between the activity level $\lambda$ and the model parameters $\hat{\alpha}$ and $\hat{\beta}$.}
    \label{fig2:alphabeta}
    \vspace*{-14pt}
\end{figure}

We thus investigate the existence of a connection between the level of activity of subscribers and the values of $\alpha$ and $\beta$ in (\ref{eq:regression}) that regulate the relationship between presence and population.
We do not have access to the real values, but to their estimations $\hat{\alpha}$ and $\hat{\beta}$.
We thus collect data in all cities that refer to the overnight period, i.e., from midnight to 8 am: in this period the ISTAT census information can be still considered a reliable ground truth, as most people will be at home.
We then plot a scatterplot of the activity level $\lambda$ versus the regression parameters $\hat{\alpha}$ and $\hat{\beta}$ obtained in these scenarios from (\ref{eq:estimreg}). The results are depicted in Fig.\,\ref{fig2:alphabeta}.

We find a striking linear relationship between $\lambda$ and both parameters. The coefficients of the linear models are indicated in the plots.
This result allows drawing a unifying multivariate model that links the population density to both the subscriber presence and the subscriber activity level.
We can then refine our estimation model as:
\begin{equation}
\hat{\rho}_i(\lambda_i,\sigma_i) =
(\hat{a}_\alpha \lambda_i+\hat{b}_\alpha) \cdot
\sigma_i^{(\hat{a}_\beta \lambda_i + \hat{b}_\beta)}.
\label{eq:dynamic}
\end{equation}

An important consideration is that (\ref{eq:dynamic}) substantiates the need for the power relationship between presence and population density in (\ref{eq:regression}) for a correct modeling of dynamic populations.
Also, the new parameters $\hat{a}_\alpha$, $\hat{b}_\alpha$, $\hat{a}_\beta$, $\hat{b}_\beta$ are valid for all scenarios, and are consistent across different times of the day.
We thus consider that the model in (\ref{eq:dynamic}) can be reliably used for the estimation of dynamic populations, given than the time series of the subscriber presence $\sigma_i$ and of the subscriber activity level $\lambda_i$ are available from mobile network traffic metadata.
		
%

\subsection{A case study in Milan}
\label{sub:dyn-milan}

\begin{figure}[tb]
	\centering
    \includegraphics[width=1\linewidth]{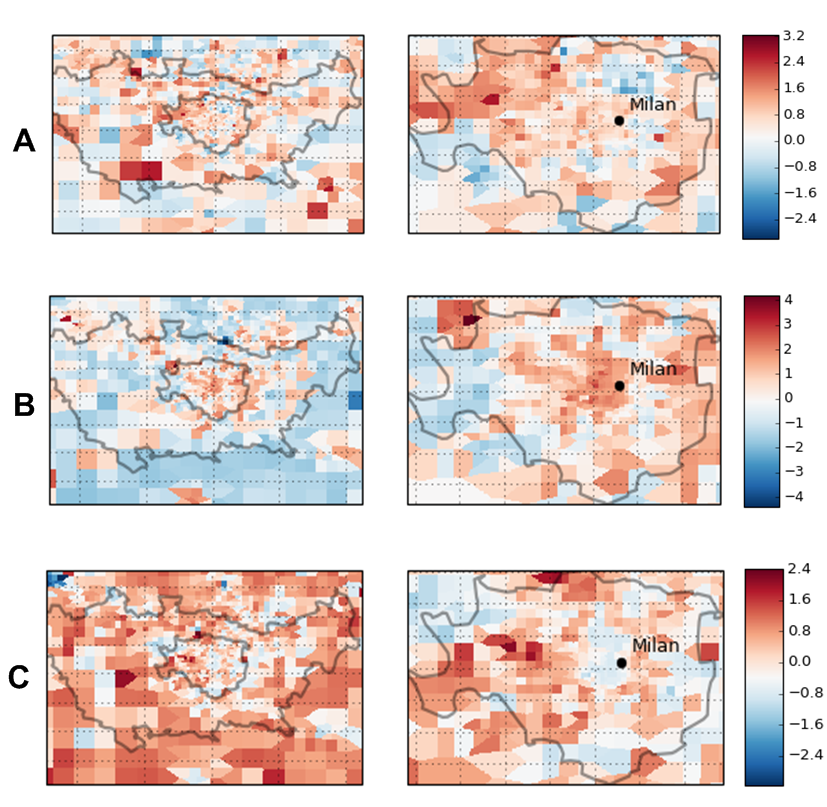}
    \caption{Dynamic distribution of Milan population. A) April 15 at noon. B) April 22 at 5 pm. C) April 19 at 10 pm. Left plots show the whole conurbation, right ones the city only. Figure best viewed in colors.}
    \label{fig1:density}
    \vspace*{-10pt}
\end{figure}

We provide a proof-of-concept of exploitation of the model in (\ref{eq:dynamic}), considering the case of Milan. Fig.\,\ref{fig1:density} illustrates the dynamic distribution of the population in Milan and its suburbs, as inferred from our model, at three sample time instants: April 15, 2015, at noon (A), April 22, 2015, at 5 pm (B) and April 19, 2015, at 10 pm (C). In each plot, colors denote the variation of population during the corresponding hour: from high in-flow of individuals moving to the cell (red) to high high out-flow of individuals leaving the cell (blue). There are also neutral cells where the population density does not vary during the considered time period (white). 

The first two times refer to subsequent Wednesdays, at inherently different hours of the day. We observe that at noon the population tends to move towards the city outskirts (A, left plot), going back home for lunch. An opposite trend is observed at 5 pm, when there is a significant in-flow towards the city center or commercial areas (B, right plot).
April 19 is a Saturday, and our estimate of the dynamic population captures the movement of people towards residential areas outside Milan (C, left plot) as well as towards nightlife areas in the city (C, right plot). Interestingly, the model conveys well the crowd attracted by an important football match that took place that day at the main city stadium (C, right plot, dark red zone towards the West).
Although we cannot compare these population variations against a proper ground truth, we remark that the model entails very reasonable behaviors that match well those known to characterize the movements of inhabitants of the Milan conurbation.

%

\section{Conclusions}
\label{sec:CONCLUSION}

We introduced a novel approach to the estimation of population based on existing relationship between mobile activity volume and population density. Our solution builds exclusively on metadata collected by mobile network operators.
Our results demonstrate how our model allows for a reliable representation of static populations across different cities. It also outperforms previous proposals in the literature, thanks to the use of more appropriate metadata as well as proper data filtering based on time and land use.

%
%

\section*{Acknowledgements}

The authors would like to thank Angelo Furno for his help with extracting land use information. This work was supported by the French National Research Agency under grant ANR-13-INFR-0005 ABCD, by the EU FP7 ERA-NET program under grant CHIST-ERA-2012 MACACO, and by BPI-France through the FUI FluidTracks project.

%
%

\bibliographystyle{IEEEtran}  
\bibliography{biblio}

\begin{thebibliography}{10}
\providecommand{\url}[1]{#1}
\csname url@samestyle\endcsname
\providecommand{\newblock}{\relax}
\providecommand{\bibinfo}[2]{#2}
\providecommand{\BIBentrySTDinterwordspacing}{\spaceskip=0pt\relax}
\providecommand{\BIBentryALTinterwordstretchfactor}{4}
\providecommand{\BIBentryALTinterwordspacing}{\spaceskip=\fontdimen2\font plus
\BIBentryALTinterwordstretchfactor\fontdimen3\font minus
  \fontdimen4\font\relax}
\providecommand{\BIBforeignlanguage}[2]{{%
\expandafter\ifx\csname l@#1\endcsname\relax
\typeout{** WARNING: IEEEtran.bst: No hyphenation pattern has been}%
\typeout{** loaded for the language `#1'. Using the pattern for}%
\typeout{** the default language instead.}%
\else
\language=\csname l@#1\endcsname
\fi
#2}}
\providecommand{\BIBdecl}{\relax}
\BIBdecl

\bibitem{song2010limits}
C.~Song, Z.~Qu, N.~Blumm, and A.-L. Barab{\'a}si, ``Limits of predictability in
  human mobility,'' \emph{Science}, vol. 327, no. 5968, pp. 1018--1021, 2010.

\bibitem{bagrow2011collective}
J.~P. Bagrow, D.~Wang, and A.-L. Barabasi, ``Collective response of human
  populations to large-scale emergencies,'' \emph{PloS one}, vol.~6, no.~3, p.
  e17680, 2011.

\bibitem{bajardi2011human}
P.~Bajardi, C.~Poletto, J.~J. Ramasco, M.~Tizzoni, V.~Colizza, and
  A.~Vespignani, ``Human mobility networks, travel restrictions, and the global
  spread of 2009 h1n1 pandemic,'' \emph{PloS one}, vol.~6, no.~1, p. e16591,
  2011.

\bibitem{yang14}
Y.~Yang, C.~Herrera, N.~Eagle, and M.~C. Gonz\'alez, ``Limits of predictability
  in commuting flows in the absence of data for calibration,'' \emph{Nature
  Scientific Reports}, vol.~4, no. 5662, 2014.

\bibitem{caceres12}
N.~Caceres, L.~Romero, F.~Benitez, and J.~Castillo, ``Traffic flow estimation
  models using cellular phone data,'' \emph{IEEE Transactions on Intelligent
  Transportation Systems}, vol.~13, no.~3, 2012.

\bibitem{naboulsi16survey}
D.~Naboulsi, M.~Fiore, R.~Stanica, and S.~Ribot, ``Large-scale mobile traffic
  analysis: a survey,'' \emph{IEEE Communications Surveys and Tutorials},
  vol.~18, no.~1, 2016.

\bibitem{krings2009urban}
G.~Krings, F.~Calabrese, C.~Ratti, and V.~D. Blondel, ``Urban gravity: a model
  for inter-city telecommunication flows,'' \emph{Journal of Statistical
  Mechanics: Theory and Experiment}, vol. 2009, no.~07, p. L07003, 2009.

\bibitem{douglass2015high}
R.~W. Douglass, D.~A. Meyer, M.~Ram, D.~Rideout, and D.~Song, ``High resolution
  population estimates from telecommunications data,'' \emph{EPJ Data Science},
  vol.~4, no.~1, pp. 1--13, 2015.

\bibitem{deville2014dynamic}
P.~Deville, C.~Linard, S.~Martin, M.~Gilbert, F.~R. Stevens, A.~E. Gaughan,
  V.~D. Blondel, and A.~J. Tatem, ``Dynamic population mapping using mobile
  phone data,'' \emph{Proceedings of the National Academy of Sciences}, vol.
  111, no.~45, pp. 15\,888--15\,893, 2014.

\bibitem{csaji2013exploring}
B.~C. Cs{\'a}ji, A.~Browet, V.~A. Traag, J.-C. Delvenne, E.~Huens,
  P.~Van~Dooren, Z.~Smoreda, and V.~D. Blondel, ``Exploring the mobility of
  mobile phone users,'' \emph{Physica A: Statistical Mechanics and its
  Applications}, vol. 392, no.~6, pp. 1459--1473, 2013.

\bibitem{bekhor2013evaluating}
S.~Bekhor, Y.~Cohen, and C.~Solomon, ``Evaluating long-distance travel patterns
  in israel by tracking cellular phone positions,'' \emph{Journal of Advanced
  Transportation}, vol.~47, no.~4, pp. 435--446, 2013.

\bibitem{calabrese2011estimating}
F.~Calabrese, G.~Di~Lorenzo, L.~Liu, and C.~Ratti, ``Estimating
  origin-destination flows using mobile phone location data,'' \emph{IEEE
  Pervasive Computing}, vol.~10, no.~4, pp. 0036--44, 2011.

\bibitem{ratti06}
C.~Ratti, R.~Pulselli, S.~Williams, and D.~Frenchman, ``Mobile land- scapes:
  Using location data from cell-phones for urban analysis,'' \emph{Environment
  and Planning B Planning and Design}, vol.~33, no.~5, 2006.

\bibitem{kang2012towards}
C.~Kang, Y.~Liu, X.~Ma, and L.~Wu, ``Towards estimating urban population
  distributions from mobile call data,'' \emph{Journal of Urban Technology},
  vol.~19, no.~4, pp. 3--21, 2012.

\bibitem{furno2015comparative}
A.~Furno, R.~Stanica, and M.~Fiore, ``A comparative evaluation of urban fabric
  detection techniques based on mobile traffic data,'' in \emph{IEEE/ACM
  ASONAM}, 2015, pp. 689--696.

\end{thebibliography}
\end{document}